\documentclass[prb,twocolumn,showpacs, superscriptaddress]{revtex4}
\usepackage{graphicx}
\usepackage{amsmath}
\usepackage{amsxtra}
\usepackage{amssymb}
\usepackage{dcolumn}
\usepackage{float}
\usepackage{bm}
\usepackage[breaklinks=true,colorlinks,citecolor=blue,linkcolor=blue,urlcolor=blue]{hyperref}

\def\be{\begin{equation}}
\def\ee{\end{equation}}
\def\bea{\begin{eqnarray}}
\def\eea{\end{eqnarray}}

\begin{document}
\author{Rashi Sachdeva}
\affiliation{Department of Physics, Indian Institute of Technology Kanpur, Kanpur 208016, India}
\affiliation{Quantum systems unit, Okinawa institute of Science and Technology graduate university, Okinawa, Japan}
\title{Plasmon modes of a massive Dirac plasma, and their superlattices}
\author{Anmol Thakur}
\affiliation{Department of Physics, Indian Institute of Technology Kanpur, Kanpur 208016, India}
\author{Giovanni Vignale}
\affiliation{Department of Physics, University of Missouri, Columbia, Missouri 65211, USA}
\author{Amit Agarwal}
\email{amitag@iitk.ac.in}
\affiliation{Department of Physics, Indian Institute of Technology Kanpur, Kanpur 208016, India}

\date{\today}

\begin{abstract}
We explore the collective density oscillations of a collection of charged massive Dirac particles, in one, two and three dimensions and their one dimensional superlattice. We calculate the long wavelength limit of the dynamical polarization function analytically,  and use the random phase approximation to obtain the plasmon dispersion. 
The density dependence of the long wavelength plasmon frequency in massive Dirac systems is found to be different compared to systems with parabolic, and gapless Dirac dispersion.  We also calculate the long wavelength plasmon dispersion of a 1d metamaterial made from 1d and 2d massive Dirac plasma. Our analytical results will be useful for exploring the use of massive Dirac materials as electrostatically tunable plasmonic metamaterials and can be experimentally verified by infrared spectroscopy as  in the case of graphene [L. Ju. {\it et. al.}, Nat. Nanotechnol.  6, 630 (2011)]. 
\end{abstract}

\maketitle
  
\section{Introduction}

The collective density oscillations of electrons liquids, i.e., plasmons,  offer a powerful tool for exploring electron-electron interaction effects in various systems  \cite{Pines, Giuliani_and_Vignale} and have also motivated several potential applications in optical metamaterials, nanophotonic lasers and amplifiers, biochemical sensing, and antennas transmitting and receiving light signals at the nanoscale \cite{Maier, QP}.
The collective modes of ordinary (Schr\"odinger) electrons with parabolic dispersion \cite{Pines, Giuliani_and_Vignale, Ando} including spin-orbit coupling \cite{ag1} and spin polarization\cite{ag2} have been extensively studied  in metals and doped semiconductors. Since the discovery of graphene, there has been huge interest in plasmons of Dirac materials and particularly in graphene\cite{GTP, Yan, Yoon, Yan2, Polini_rev, Stauber,Fei} as it offers a tunable plasmon spectrum via electrostatic control of its carrier concentration, and higher plasmon lifetimes due to high mobility. 

There have been several studies on plasmons in gapless  two dimensional (2d) and three dimensional (3d) Dirac systems in the context of graphene \cite{Guinea_NJP2007, SDS1, Polini1, SDS2, Saeed}, topological insulators \cite{RaghuPRL2009, zhang_ijmp_2013, Pietro}, Weyl semimetals \cite{xiao_arxiv_2014}, and in gapped 2d massive Dirac systems \cite{Pyatkovskiy} in the context of buckled honeycomb structures such as silicene \cite{Tabert, Jianhui, Duppen}. In addition to this plasmons in 
periodic arrays of parabolic systems \cite{SDSp1, SDSp2, Guinea_PRL, Rodin}, and massless Dirac plasma layers \cite{SDS12, Peeters, Rossi} have also been investigated. A metamaterial made up of periodic graphene micro-ribbon arrays was used in Ref.~[\onlinecite{GTP}] to demonstrate tunable terahertz plasmon excitations in graphene. Comparatively  massive Dirac system in various dimensions, and its multilayers/superlattice have been relatively less explored, and consequently are the subject of this article.

In this article we study the plasmon frequency, and its density dependence for a  massive Dirac plasma (MDP) interacting via the long ranged Coulomb interaction, in one-, two- and three dimensions. Additionally we also calculate the plasmon dispersion for metamaterials made of 1d nano-ribbons and 2d layers of  MDP.  The gapless Dirac systems were studied in Ref.~[\onlinecite{SDS2}], which serves as a check of all of our calculations in the limit of the vanishing band gap. We find that while the long-wavelength plasmon frequency in MDP is essentially quantum mechanical in nature with $1/\sqrt{\hbar}$ appearing explicitly in the plasmon dispersion as in the case of gapless Dirac plasma (GDP), the scaling of the plasmon frequency with density is different for MDP, GDP and parabolic systems. 
Note that in systems with non-relativistic parabolic dispersion, the long wavelength plasmon frequency is `classical' and quantum corrections (arising from the self energy and vertex corrections in the polarization function) show up only in higher order terms. 
 The aim of this work is to illustrate the key differences between the density dependences of plasmon dispersions in one-, two- and three-dimensional systems
which arise due to the relativistic (Dirac) or non-relativistic (Schr\"odinger) nature of the electrons and  due to the presence of a finite gap in MDP.

This article is organized as follows: In Sec.~\ref{polfuncsec}, we introduce the random phase approximation (RPA) `recipe' for calculating the plasmon frequency and  
explicitly calculate the long wavelength limit of the dynamical polarization function for MDP, GDP and parabolic dispersion systems. 
This allows us to obtain and discuss similarities and differences in the long wavelength plasmon frequencies  in Sec.\ref{plasmonlw}.  Next we consider the plasmons arising in a periodic array of MDP 
nano-ribbons and layers in Sec.~\ref{plasarraysec} and compare the results with GDP and parabolic dispersion systems. Finally, in Sec.~\ref{summarysec}, we summarize our findings.  

\section{Polarization function}\label{polfuncsec}
Within the RPA, the collective plasmon modes of an electron system emerge as poles of the density density response functions (also called polarization function or the Lindhard function) and coincide with the zeros of the complex longitudinal ``dielectric function" $\epsilon(q,\omega)$, {\it i.e.} 
\be \label{eq:condition}
\epsilon(q, \omega) = 1 - v_q \Pi(q, \omega) =  0~, 
\ee
where $v_q$ is the  Fourier transform of the Coulomb interaction, and $\Pi$ is the total non-interacting polarizability of the system. 
The Fourier transform of the Coulomb interaction $v (r) = e^2/(\kappa r)$, in the appropriate $d$-dimensional space is given by 
\begin{subequations} \label{vq}
\begin{align}
      v_q &= \frac{4 \pi e^2}{\kappa q^2} \quad \quad d = 3~,  \\
       & =  \frac{2 \pi e^2}{\kappa q} \quad \quad d = 2~, \\
	& =  \frac{2 e^2}{\kappa}K_{0}(qa) \quad   d = 1~,
\end{align}
\end{subequations}
where $\kappa$ is the background material dependent dielectric constant, and $K_0$ denotes the zeroth order modified Bessel function of the second kind. Note that in one dimension, the length scale $a$ characterizes   the lateral confinement size (say radius of the 1d ribbon), and $v_q\approx -2 e^2 \ln(qa)/\kappa$ for $qa \ll1$, while $v_q = e^2/(\kappa q^2 a^2)$ for $q a \gg 1$.  

The polarization function for the massive Dirac material is given by
\be \label{Lindhard}
\Pi(q,\omega)=\frac{g_s g_v}{L^d} \sum_{{\bf k}, \lambda,\lambda'}F_{\lambda,\lambda'}({\bf k}, {\bf k'})~ \frac{n_{\rm F}(\lambda E_{\textbf{k}})-n_{\rm F}(\lambda'E_{{\bf k'}})}{\hbar\omega + \lambda E_{\textbf{k}} -  \lambda'E_{ {\bf k'}} + i\eta }~,
\ee
where ${\bf k'} = {\bf k} + \bf {q}$, $\lambda, \lambda' = \pm 1$ denotes the conduction (particle) and valence (hole) bands,  $E_{\textbf{k}}= \hbar v_{\rm F}| \sqrt{k^2+(\Delta/\hbar v_{\rm F})^2}|$  with $2 \Delta$ being the energy gap, $n_{\rm F} (x)$ is the Fermi function and $2 F_{\lambda,\lambda'}({\bf k}, {\bf k'}) =1+\lambda\lambda' [\textbf{k}\cdot{\bf k'}+\tilde{\Delta}^2]/(\tilde{E}_\textbf{k} \tilde{E}_{\textbf k'})$ is the overlap function, with $\tilde{x} \equiv x/\hbar v_{\rm F}$. The factor  $g_s~(=2)$ is the spin degeneracy factor and $g_v$ is the valley (or pseudo spin) degeneracy factor (e.g. $g_v = 2$ for graphene and other Dirac materials with honeycomb lattice structure). Given the general relation $\Pi(q, -\omega) = \Pi(q, \omega)^{*}$, 
and the fact that the polarization function depends only on the absolute value of the Fermi energy $\varepsilon_{\rm F}$,  we only present the results for $\varepsilon_{\rm F} > 0$ and $\omega >0$. Furthermore, we work at zero temperature so that the Fermi functions  can be replaced by Heaviside step functions, i.e., $n_{\rm F}(x) = \Theta(\varepsilon_{\rm F}-x)$.

Depending upon the placement of the Fermi energy $\varepsilon_{\rm F}$, we can split our polarization  function into two parts, namely the intrinsic  ($\varepsilon_{\rm F} < \Delta$) and extrinsic ($\varepsilon_{\rm F} > \Delta$) polarization:
\bea
 \Pi~(q,\omega)
&=&-\chi_{\infty}^-(q,\omega)+ \underbrace{\chi_{\varepsilon_{\rm F}}^-(q,\omega)+ \chi_{\varepsilon_{\rm F}}^+(q,\omega)} \nonumber\\
&=& \Pi_0(q,\omega)+\theta(\varepsilon_{\rm F}-\Delta)~\Pi_1(q,\omega)\label{totalpol}~,
\eea
where 
\bea & & \chi_{D}^\pm(q,\omega)=-\frac{g_s g_v}{(2\pi)^d} \int d^d
k ~\Theta(D^2-\Delta^2-k^2)\\
& & \times\left(1\pm \frac{\textbf{k}\cdot\textbf{k}'+\tilde{\Delta}^2}{\tilde{E}_{\textbf{k}}~ \tilde{E}_{\textbf{k}'}}\right)
\left[\frac{E_{\textbf{k}}\mp E_{\textbf{k}'}}{(\hbar\omega+i\eta)^2-(E_{\textbf{k}}\mp E_{\textbf{k}'})^2}\right]~. \nonumber
\eea
Here the upper and lower signs correspond to intraband and interband electron-hole transitions respectively and the parameter $D$ defines the integration limits via the $\Theta$ function. Since we are interested in the long wavelength ($q\to 0$) plasmon dispersion, we evaluate Eq.~\eqref{Lindhard} in the dynamical limit ($q \to 0$ first and then $\omega \to 0$), to lowest order in $q^2/\omega^2$, 
just above the intra-band particle hole continuum.

We mention at the outset that we will use the superscript $(\rm p)$, $(\rm g)$ and $(\rm m)$ to refer to systems with parabolic, gapless (or massless) Dirac, and massive Dirac systems, respectively. Note  that the electronic density for any d-dimensional system, in terms of its Fermi wavevector is given by 
\be n_d = g_s g_v \frac{\pi^{d/2} {k_{\rm F}}^d}{2^d \pi^d \Gamma(1+d/2)}~,\ee
where $\Gamma(x)$ is the Gamma function. 
 However the Fermi wavevectors for parabolic, massive  Dirac, and gapless Dirac systems are expressed differently  in terms of the Fermi energy and  are given by $k_{\rm F} = \sqrt{2 m \varepsilon_{\rm F}}/\hbar$, $k_{\rm F} = \sqrt{\varepsilon_{\rm F}^2-\Delta^2}/\hbar v_{\rm F}$ and $k_{\rm F} = \varepsilon_{\rm F}/\hbar v_{\rm F}$, respectively.

For systems with parabolic dispersion ($E_{\textbf{k}} = \hbar^2 k^2/2 m_{\rm p}$),  Eq.~\eqref{Lindhard} can be evaluated in the dynamical long wavelength limit, upto leading order in $q$ just above the particle-hole continuum to obtain
\be \label{eq:Pi_parabolic}
\Pi^{(\rm p)} (q, \omega) \approx \frac{n_d}{m_{\rm p}} \frac{q^2}{\omega^2} + {\cal O}\left(\frac{q^4}{\omega^4}\right)~. 
\ee
For massive Dirac systems, $E_{\bf k} =\hbar v_{\rm F}\sqrt{ k^2 + \tilde\Delta^2} $ (where $\tilde{\Delta} = \Delta/\hbar v_{\rm F}$) in all dimensions, and the dynamical long wavelength limit of the Lindhard function, just above the particle-hole continuum, is given by
\bea \label{eq:Pi_massive}
\Pi^{(\rm m)} &\approx&  \frac{ g_s g_v v_{\rm F}}{\hbar (2 \pi)^d} \frac{ \pi^{d/2}}{\Gamma(1+d/2)} \frac{k_{\rm F}^d}{\sqrt{k_{\rm F}^2 + {\tilde \Delta}^2}} \frac{q^2}{\omega^2}+ {\cal O}\left(\frac{q^4}{\omega^4}\right)~. \nonumber \\ \eea
For  massless Dirac systems such as graphene, $\Delta \to 0$, and $E_{\bf k} = \hbar v_{\rm F} k$ in all dimensions,  and 
Eq.~\eqref{eq:Pi_massive} reduces to 
\bea  \label{eq:gapless}
\Pi^{(\rm g)} (q, \omega) &\approx& \frac{ g_s g_v v_{\rm F} k_{\rm F}^{d-1}}{\hbar (2 \pi)^d} \frac{ \pi^{d/2}}{\Gamma(1+ d/2)} \frac{q^2}{\omega^2} +{\cal O}\left(\frac{q^4}{\omega^4}\right), \nonumber \\~\eea
which is consistent with Eq.~(6) of Ref.~[\onlinecite{SDS2}]. We emphasize here that even though the density dependence of the long wavelength limit of the polarization function for  massless and massive Dirac systems is different from that of the  parabolic systems, they can be rewritten in the same form as Eq.~\eqref{Lindhard}, 
\bea \label{eq:similar}
 \Pi^{(\rm m,g)} (q, \omega) &\approx & \frac{ n_d}{\varepsilon_{\rm F}/v_{F}^2} \frac{q^2}{\omega^2}+ {\cal O}\left(\frac{q^4}{\omega^4}\right).
~\eea
Note the similarity between Eq.~\eqref{eq:Pi_parabolic} and Eq.~\eqref{eq:similar}.  This prompts the following mapping: band mass in parabolic systems,  $m_{\rm p}\to m_{\rm d} \equiv \varepsilon_{\rm F}/ v_{\rm F}^2$, density dependent effective Dirac mass in massive as well as massless Dirac systems (to be distinguished from the band gap $\Delta$ which is also  occasionally referred to as mass ). As a natural consequence, this correspondence manifests itself in all of the subsequent calculations.  

Physically the Dirac mass is a dynamical collective mass and is essential to explain inertial acceleration of the Dirac plasma under application of an external electric field. In fact it has been insightfully defined as `plasmon mass' in the context of graphene \cite{Saeed}, and has also been recently  measured in graphene \cite{Yoon}.
Note that $m_{\rm d} =\varepsilon_{\rm F}/v_{\rm F}^2$ is also the cyclotron effective mass ($m_{\rm c}$) for Dirac systems \cite{Wright}, which is typically defined as $2 \pi m_c = \hbar^2 d S(\varepsilon)/d\varepsilon $, where $ S(\varepsilon) = \pi (\varepsilon^2 - \Delta^2)/ \hbar^2 v_{\rm F}^2 $ denotes the area of a closed cyclotron orbit of massive Dirac electrons with energy $\varepsilon$. For a system with a parabolic dispersion, the band mass, plasmon mass and the cyclotron mass are identical, $m_{\rm p} = m_{\rm c}$.

Having obtained the long wavelength limit of the dynamical polarization function, we now proceed to calculate the long wavelength limit of the plasmon dispersion in the next section. 

\section{Plasmon Dispersion}\label{plasmonlw}
Using Eq.~\eqref{eq:Pi_parabolic} and Eq.~\eqref{vq} in Eq.~\eqref{eq:condition}, the well known long wavelength plasmon dispersion for systems with parabolic dispersion \cite{Ando, Giuliani_and_Vignale}, in one-, two- and three dimensions can be easily obtained  to be
\begin{subequations}
\label{wp}
\begin{align}
\omega^{({\rm p})}_1 &=\sqrt{ \frac{2e^2 n_1}{\kappa m_{\rm p}}} q\sqrt{K_0(qa)}+ {\cal O}(q^3)~, \label{wp1}\\
\omega^{({\rm p})}_2 &= \sqrt{ \frac{2\pi e^2 n_2}{\kappa m_{\rm p}}} q^{1/2}+ {\cal O}(q^{3/2}) ~, \label{wp2}\\
\omega^{({\rm p})}_3 &=  \sqrt{ \frac{4\pi e^2 n_3}{\kappa m_{\rm p}}}+ {\cal O}(q^2) ~. \label{wp3}
\end{align}
\end{subequations}
An important point to note here is that for the first term in Eq.~\eqref{wp}, we can substitute $m_{\rm p} \to m$, i.e., replace the effective band mass by  the classical mass of the particle and the quantum mechanical plasmon dispersion takes precisely the same form as that of classical density oscillations in an electron liquid \cite{Pines, Ando, Giuliani_and_Vignale}. Physically this is a direct consequence of the fact that the long wavelength plasmons involve the motion of the entire plasma, and to lowest order it does not depend on the complex exchange and correlation
effects that dress the motion of an individual electron. It should be emphasized that  no such classical analog exists for Dirac systems, and the plasmon dispersion in Dirac systems is intrinsically quantum mechanical in nature\cite{SDS2}. Note, however, that the higher order correction terms in Eq.~\eqref{wp} are fully quantum mechanical and $\hbar$ explicitly appears in them.

For systems with massive Dirac dispersion, using Eqs.~\eqref{eq:Pi_massive} and~\eqref{vq} in Eq.~\eqref{eq:condition},  the $q\to 0$ limit of the plasmon dispersion is given by 
\begin{subequations}
\label{wm}
\begin{align}
\omega^{({\rm m})}_1 &=  \sqrt{ \frac{2 g e^2 v_{\rm F}}{ \hbar \kappa \pi}} q\sqrt{K_0 (qa)}\frac{(\varepsilon_{\rm F}^2-\Delta^2)^{1/4}}{\varepsilon_{\rm F}^{1/2}}+ {\cal O}(q^3),\nonumber  \\
 \label{wm1}\\ 
\omega^{({\rm m})}_2 &= \sqrt{ \frac{g e^2 }{2 \kappa\hbar^2}} \sqrt{\frac{\varepsilon_{\rm F}^2-\Delta^2}{\varepsilon_{\rm F}}}q^{1/2}+ {\cal O}(q^{3/2}) \label{wm2}~,\\ 
\omega^{({\rm m})}_3 &=   \sqrt{ \frac{2 g e^2 }{3\pi \kappa\hbar^3 v_{\rm F}}}\frac{(\varepsilon_{\rm F}^2-\Delta^2)^{3/4}}{\varepsilon_{\rm F}^{1/2}} + {\cal O}(q^{2})~,
 \label{wm3}
 \end{align}
\end{subequations}
where we have defined $g \equiv g_s g_v$. 
Note that 2d massive Dirac plasma was also studied in Ref.~[\onlinecite{Pyatkovskiy}], which reported an expression similar to Eq.~\eqref{wm2}.
For the limiting case of $\Delta \to 0$, Eq.~\eqref{wm} leads to, 
\begin{subequations}
\label{wg}
\begin{align}
\omega^{({\rm g})}_1 &=  \sqrt{ \frac{2g e^2 v_{\rm F}}{\hbar \kappa \pi}} q\sqrt{K_0 (qa)}+ {\cal O}(q^3)~,
 \label{wg1}\\
\omega^{({\rm g})}_2 &=   \sqrt{ \frac{g e^2 \varepsilon_{\rm F}}{2 \kappa\hbar^2}} q^{1/2}+ {\cal O}(q^{3/2})
 ~, \label{wg2} \\
\omega^{({\rm g})}_3 &=  \sqrt{ \frac{2 g e^2 \varepsilon_{\rm F}^2}{3\pi \kappa\hbar^3 v_{\rm F}}} + {\cal O}(q^{2})
 ~. \label{wg3}
 \end{align}
\end{subequations}
Equation~\eqref{wg} reproduces the results for gapless Dirac plasma reported in Ref.~[\onlinecite{SDS2}]. 
One important similarity between Eqs.~\eqref{wp},~\eqref{wm} and~\eqref{wg}, is the same functional dependence of the plasmon frequency on the wave vector $q$. This is direct consequence of  the physical requirement that the long wavelength plasmon dispersion must satisfy particle conservation (or continuity equation). One important difference between parabolic and Dirac systems is that while the long wavelength limit of plasmon dispersion in parabolic systems is essentially `classical' in nature, the plasmon dispersion in GDP and MDP is essentially quantum mechanical, as evidenced by the explicit appearance of $\hbar$ in Eqs.~\eqref{wm} and~\eqref{wg}.

The long wavelength dependence of the dynamical polarization function is the same for parabolic systems, GDP and MDP: $\Pi \propto q^2/\omega^2$,  however the proportionality constant has a different density dependence for various systems.  For parabolic systems $ \Pi^{(\rm p)} \propto n_d$, for gapless Dirac systems $\Pi^{(\rm g)} \propto n_d^{1-1/d}$, and for massive Dirac systems $\Pi^{(\rm m)} \propto n_d/(n_d^{2/d}+\alpha_d \tilde\Delta^{2})^{1/2}$, where $\alpha_d = (g/\pi)^2,~g/4 \pi,~ (g/6 \pi^2)^{2/3} $ in 1, 2, and 3 dimensions respectively.
As a consequence, the density dependence of the long wavelength plasmon frequency for MDP is completely different  compared to GDP and parabolic dispersion systems.
As seen from Eq. ~\eqref{wp}, the plasmon frequency for parabolic dispersion system is proportional to $\sqrt{n_d}$ in all dimensions. However, for GDP the plasmon dispersion follows $\omega_d^{(g)} \propto \sqrt{n_{d}}/n_{d}^{1/2d} $ behavior and for the one dimensional case, the plasmon mode is completely independent of the density. For the MDP case, the density dependence completely changes due to the presence of gap and takes the form 
$\omega^{({\rm m})}_{d}  \propto  \sqrt{n_d}/(n_d^{2/d}+\alpha_d \tilde\Delta^{2})^{1/4}$,  
where $\alpha_d = (g/\pi)^2,~g/4 \pi,~ (g/6 \pi^2)^{2/3} $, in 1, 2, and 3 dimensions respectively.
Note that in one dimension, the plasmon frequency in GDP is independent of the density whereas for MDP, 
the plasmon frequency of MDP has an explicit dependence on $n_{1}$ as evident from Eq. ~\eqref{wm1}.

Similar to the case of the long wavelength polarization function in Eq.~\eqref{eq:similar}, the plasmon frequencies  in the $q \to 0$  limit in GDP and MDP can also be rewritten in the form similar to Eq.~\eqref{wp} for parabolic systems, 
\begin{subequations}
\begin{align}
\omega^{({\rm m,g})}_1 &= \sqrt{ \frac{2e^2 n_1}{\kappa m_{\rm d}}} q\sqrt{K_0 (qa)}+ {\cal O}(q^3)\label{msim1} ~,\\
\omega^{({\rm m,g})}_2 &=\sqrt{ \frac{2\pi e^2 n_2}{\kappa m_{\rm d} }} q^{1/2}+ {\cal O}(q^{3/2})~\label{msim2}~,\\
\omega^{({\rm m,g})}_3 &=\sqrt{ \frac{4\pi e^2 n_3}{\kappa m_{\rm d} }}+ {\cal O}(q^2)~.\label{msim3}~
\end{align}
\end{subequations}
We emphasize again that even though the density dependence of plasmon frequencies in long wavelength limit in the GDP and MDP  is different from that of the parabolic systems, the frequencies can be rewritten in the same form using the density dependent effective Dirac mass (or equivalently the cyclotron mass). However, we note that this simplicity is deceptive, since for parabolic systems the band mass $m_{\rm p}$ and density $n_d$ can also be treated as classical independent variables, but for Dirac systems $m_{\rm d}$ is density dependent via the Fermi energy in a purely quantum mechanical way.

\section{Plasmons in metamaterials made of massive Dirac plasma, ribbon and layer, arrays}\label{plasarraysec}
In this section we consider collective density excitations in metamaterials (periodic arrays) of massive Dirac plasma systems. In particular we consider a stacking of identical 1d massive Dirac plasma nano-ribbons (or quantum wires) placed parallel to each other in a plane, and a periodic array of parallel 2d massive Dirac plasma sheets. 
Similar systems made of parabolic dispersion and gapless Dirac plasma have been theoretically investigated earlier \cite{SDSp1, SDSp2, SDS2, Peeters,Guinea_PRL} and experimentally demonstrated for graphene \cite{GTP, Yan, Yan2} . To describe the collective modes of such superstructures, we need to include the inter-ribbon or inter-layer Coulomb interactions, which leads to a coupling of all the layers due to the long range nature of Coulomb interactions. Assuming no wave function overlap between any consecutive layers or nano-ribbons, the collective modes of such superlattices, within RPA,  are given by the zeros of the determinant of the general dielectric  matrix of the superlattice, whose elements are given by  
\be \label{eq:sl}
\epsilon_{ll'}  = \delta_{ll'} - v_{ll'}(q, k)  \Pi_{l'} (q,\omega)  ,  
\ee
where $\Pi_{l} (q, \omega) = \Pi (q,\omega)$ is the bare density-density response function of each nano-ribbon or layer whose long wavelength limit is given in Eq.~\eqref{eq:Pi_massive}. In Eq.~\eqref{eq:sl},  $v_{ll'}(q, k)$ is the repulsive Coulomb interaction between the $l$ and $l'$ nano-ribbon or layer in the periodic array which is given by 
\begin{subequations}
\begin{align}
v_{ll'} &= \frac{2 \pi e^2}{\kappa q} e^{-q b | l- l'|} \quad \quad d=2~ , \\
v_{ll'} &=  \frac{2 e^2}{\kappa } \left[\delta_{ll'} K_0 (q a) + (1-\delta_{ll'})K_0 (q b |l-l'|)\right] \quad d = 1~, \nonumber\\ 
\end{align}
\end{subequations}
where $b$ is the superlattice spacing. 

\begin{figure}[t]
\includegraphics[width=0.80\linewidth]{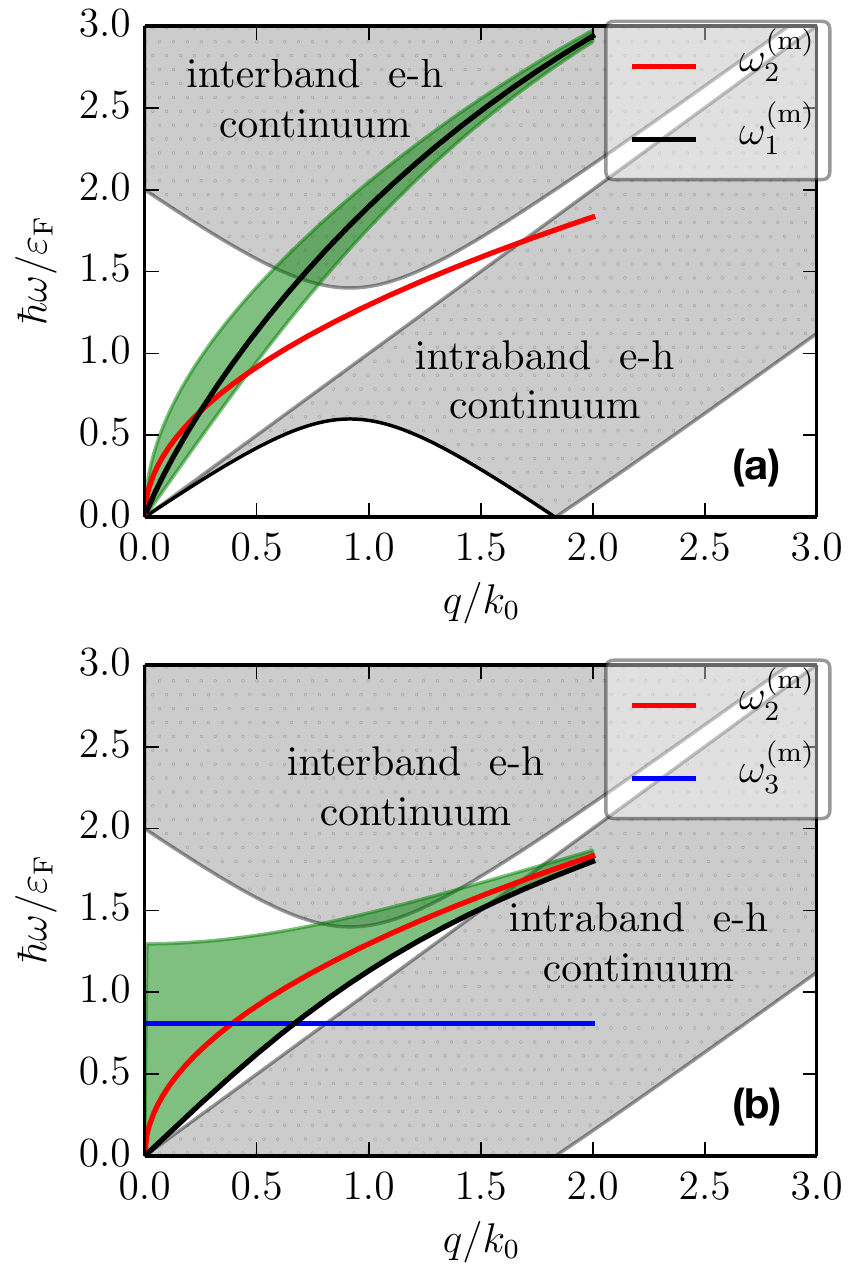}
\caption{  (a) The long wavelength plasmon dispersion for 1d and 2d MDP against the backdrop of the 1d particle-hole continuum (the grey shaded region). The green shaded region marks the plasmon band formed in the 1d superlattice of the 1d MDP [the lower boundary is for $q_y =\pi/b$, and the upper boundary is for $q_y = 0$ in  Eq.~\eqref{eq:gen}].
 (b) The long wavelength plasmon dispersion for 2d and 3d MDP, with the green shaded region marking the plasmon band formed in the 1d superlattice of the 2d MDP [the lower boundary is for $q_z =\pi/b$, and the upper boundary is for $q_z = 0$]. 
The gray shaded region in panel (b) depicts the particle-hole continuum of 2d and 3d massive plasma.  
Note that as in the parabolic case, the electron-hole continuum in 1d differs from the 2d and 3d case because there are no excitations at finite $q$ and low $\omega$ even for massive and massless Dirac fermions.
In both panels we have defined $\hbar v_{\rm F} k_0 \equiv \varepsilon_{\rm F} $ and used the following parameters: $\Delta/\varepsilon_{\rm F} = 0.4$, $ge^2/(2\hbar \kappa v_{\rm F}) = 2$,  $a k_0 = 0.25$, and $b k_0 = 2$.
\label{fig1}}
\end{figure}

Assuming a periodic boundary conditions for the 1d superlattice, the eigenvalues of the general dielectric matrix in Eq.~\eqref{eq:sl} are given 
by $1 -  v_q \Pi (q,\omega) S_d(q,k) $ where $ S_d(q,k) = v_q^{-1} \sum_{l'} v_{ll'} e^{-ik (l-l')b}$ 
is the form factor for the 1d superlattice formed from a d-dimensional plasma,  and $k$ is to be interpreted as a new wave vector arising form the periodicity of the infinite superlattice array and $|k| < \pi /b$. For definiteness we take the 1d MDP wire array to be along the $x$ axis ($q = q_x$), and $k = q_y$ to be along  the superlattice direction of the $y$ axis. For the 2d MDP layer superlattice, we consider it to lie in the $x-y$ plane and the wave vector $k = q_z$ to be along the superlattice direction --- the $z$ axis. The dimensionless form factors can now be evaluated and are given by  
\begin{subequations}
\label{eq:S}
\begin{align}
S_1 & =  1 + \frac{2}{K_0(q a)} \sum_{n = 1}^{\infty} K_0 ( n q b) \cos(n q_y b)~\label{eqS1}~, \\
S_2 &= \sum_{n = - \infty}^{\infty} e^{-q|n| b -i q_z n b} =   \frac{\sinh (qb)}{\cosh(qb) - \cos(q_z b)}~. \label{eqS2}\nonumber \\
\end{align}
\end{subequations}
Note that for $q_z =0$, as $q \to 0$, $S_2 \to q$. 
The plasmon bands for the 1d superlattice, $\omega^{(p,m,g)}_{ds}$ composed of Schr\"odinger electrons, MDP and GDP, are now explicitly given by the zeros of the  eigenvalues of the general dielectric matrix in Eq.~\eqref{eq:sl}:
\be
1 - v_q \Pi (q,\omega) S_d(q,k) = 0 ~. \label{eq:S1D}
\ee
However in the long wavelength dynamical limit ($q \to 0$), $\Pi \propto q^2/\omega^2$ for parabolic, massive Dirac and gapless Dirac systems in all dimensions, and consequently Eq.~\eqref{eq:S1D} simplifies to give, 
\be
\omega^{(p,m,g)}_{ds} = \omega^{(p,m,g)}_{d} S_d^{1/2}~,  \label{eq:gen}
\ee
where the form factor $S_{d}$ is explicitly given in Eq.~\eqref{eq:S}.
We emphasize that Eq.~\eqref{eq:gen} is very general and it describes the long wavelength plasmon dispersion for 1d  superlattice made of  parabolic, massive Dirac or gapless Dirac systems (for both ribbons and layers). The plasmon bands for a superlattice of 1d MDP nano-ribbons, and 2d MDP layers is displayed in panels (a) and (b) of Fig.~\ref{fig1}, respectively against the backdrop of the particle-hole continuum.

For ribbons and layers of parabolic systems, the superlattice plasmon dispersion at the upper band edge ($k=0$) is given by 
\begin{subequations}
\label{park}
\begin{align}
\omega^{(p)}_{1s} (q ; q_y =0) & =  \left(\frac{2\pi\tilde{n}_2 e^2 q}{\kappa m_{\rm p}}\right)^{1/2}~, \label{park01}\\
\omega^{(p)}_{2s}  (q ; q_z =0)  & = \left(\frac{4\pi\tilde{n}_3 e^2 }{\kappa m_{\rm p}}\right)^{1/2}  ~. \label{park02}
\end{align}
\end{subequations}
with $\tilde{n}_2=\frac{n_1}{b}$ and $\tilde{n}_3=\frac{n_2}{b}$. Note that in  Eq.~\eqref{park}, the d-dimensional superlattice plasmon at the band edge ($k=0$) has exactly the 
same form as the corresponding (d+1)-dimensional bulk plasmon [see Eqs.~\eqref{wp1}-\eqref{wp2}] with the effective densities being $\tilde{n}_3=n_2/b$ and $\tilde{n}_2=n_1/b$. This is consistent physically since the d-dimensional superlattice loses its discrete periodicity at the band edge ($k=0$) and effectively becomes a 
 (d+1)-dimensional system.

In the case of superlattice structures made of MDP, the $q\to 0$ plasmon dispersion at the band edge ($k=0$) is 
\begin{subequations}
\begin{align}
\omega^{(m)}_{1s} (q ; q_y =0)  & =   \sqrt{ \frac{2 g e^2 v_{\rm F} q}{\hbar \kappa b}} \frac{(\varepsilon_{\rm F}^2-\Delta^2)^{1/4}}{\varepsilon_{\rm F}^{1/2}}~,\label{massk01}\\
\omega^{(m)}_{2s} (q ; q_z =0) & = \sqrt{ \frac{g e^2 }{\hbar^2 \kappa b}} \sqrt{\frac{\varepsilon_{\rm F}^2-\Delta^2}{\varepsilon_{\rm F}}}~.
\label{massk03}
\end{align}
\end{subequations}
These in the limit $\Delta \to 0$, give the corresponding expressions for the GDP: 
\begin{subequations}
\begin{align}
\omega^{(g)}_{1s}  (q ; q_y =0)  & =  \left(\frac{2ge^2 v_{\rm F} q}{\hbar\kappa b}\right)^{1/2}~,\label{masslesk01} \\
\omega^{(g)}_{2s}  (q ; q_z =0) & =   \left(\frac{e^2 g\varepsilon_{\rm F}}{\hbar^2 \kappa b}\right)^{1/2}~.\label{masslesk03}
\end{align}
\end{subequations}
The physically appealing correspondence between the d-dimensional superlattice at the band edge and (d+1)-dimensional bulk system, does not happen for MDP as well as for GDP. From Eqs.~\eqref{massk01}-\eqref{massk03}, it is clear that at the band edge,  $\omega^{(m,g)}_{1s} (q;  q_y =0) \neq \omega^{(m,g)}_{2}$ with the intuitive substitution $\tilde{n}_2 = n_1/b$ and $\omega^{(m,g)}_{2s} (q;  q_z =0) \neq \omega^{(m,g)}_{3}$ with $\tilde{n}_3 = n_2/b$. This is a direct consequence of different density dependence of the polarization function [see Eqs. ~\eqref{eq:Pi_massive} and ~\eqref{eq:gapless}] in massive and massless Dirac systems as compared to systems with a parabolic dispersion relation [see Eq. ~\eqref{eq:Pi_parabolic}]. 
However,  the 2d superlattice plasmon dispersion would map to the corresponding plasmon dispersion for 3d massive Dirac plasma $\omega^{(m)}_{2s} (q; q_z=0, n_2) \to \omega^{(m)}_{3} (q, \tilde{n}_3)$,  if rather than the intuitive definition $\tilde{n}_3=n_2/b$, we have the following correspondence of densities in 3d and 2d massive systems, 
\be 
 \left(\frac{ \tilde{n}_{3}b}{n_2}\right)^{2}= \frac{\left(6\pi^2\tilde{n}_3 g^{-1}\right)^{2/3}+\tilde\Delta^{2}}{4\pi n_2g^{-1}+\tilde\Delta^{2}}~.\label{2dto3d}
\ee 
In the  $\Delta\to 0$ limit, Eq.~[\ref{2dto3d}]  leads to the corresponding relation  GDP, i.e., $\tilde{n}_3 =  (9 \pi g/16)^{1/4} (n_2/b^2)^{3/4} $, which was first derived in Ref.~[\onlinecite{SDS2}]. For the 1d superlattice, the gapless Dirac plasma frequency $\omega^{(g)}_{1s}$ does not  depend on the carrier density at all, and hence massive Dirac plasma frequency differs here in this aspect. For 1d massive Dirac plasma, the superlattice plasmon frequency at the band edge  would agree with the 2d massive plasma frequency, i.e., $ \omega^{(m)}_{1s}(q; q_y =0, n_1) \to \omega^{(m)}_{2} (q, \tilde{n}_2)$,  only if we have the following relation between the densities $\tilde{n}_2$ and $n_1$.
 \be
 \left(\frac{\tilde{n}_2 b}{n_1}\right)^2=\frac {4 \pi \tilde{n}_2  g+ g^2 \tilde\Delta^{2}}{\pi^2 n_1^2 + g^2 \tilde\Delta^{2}}~. \label{n2ton1}
\ee
In the  $\Delta\to 0$ limit, Eq.~[\ref{n2ton1}]  reproduces the corresponding relation  GDP, i.e., $\tilde{n}_2 = 4 g/\pi b^2  $. 

Due to the presence of gap in massive Dirac systems, the density dependence as well as the band edge plasmon at $k=0$ differs from the usual parabolic as well as the massless Dirac systems. This is because of the fact that the density dependence of the polarizability of massive Dirac systems is completely different as compared to massless Dirac and parabolic dispersion systems,  as shown in Eqs. ~\eqref{eq:Pi_massive}. 

Finally, we note that as in the case of bulk plasmons, in Eqs.~\eqref{msim1}-\eqref{msim3}, the superlattice plasmon frequencies for Dirac systems can be expressed in the same form as for systems with parabolic dispersion relation.  Expressing the numerator in Eqs.~\eqref{massk01}-\eqref{massk02} in terms of density, and the denominator in terms of the cyclotron mass of massive Dirac particles $m_{\rm d}=\varepsilon_{\rm F}/ v_{\rm F}^2$, we have 
\begin{subequations}
\begin{align}
\omega^{(m,g)}_{1s} (q ; q_y =0) &=  \left(\frac{2\pi\tilde{n}_2e^2 q}{\kappa m_{\rm d} }\right)^{1/2}~,\label{massk02}\\
\omega^{(m,g)}_{2s}  (q ; q_z =0)  &= \left(\frac{4\pi\tilde{n}_3e^2 }{\kappa  m_{\rm d}}\right)^{1/2}~. \label{massk04} 
\end{align}
\end{subequations}

\section{Summary and conclusion}\label{summarysec} 

In this article we have obtained the long wavelength plasmon frequency for massive Dirac particles  in various dimensions and their 1d superlattice, and compared it with the corresponding results for plasmons in parabolic systems and gapless Dirac systems. As expected,  factors of $1/\sqrt{\hbar}$ explicitly appear even in the leading order term in the long wavelength  plasmon dispersion of MDP and GDP, highlighting their intrinsically nonclassical and quantum nature. 

To summarize we find that the long wavelength limit of the dynamical density response function, while having the same dependence for $q$ and $\omega$, i.e.,  $\Pi \propto q^2/\omega^2$ for parabolic as well as Dirac systems, has a different dependence on density. This different density dependence also gets manifested in the long wavelength plasmon dispersion and we find that for massive Dirac systems $\omega^{({\rm m})}_{d}  \propto  \sqrt{n_d}/(n_d^{2/d}+\alpha_d \tilde\Delta^{2})^{1/4}$ while for  gapless Dirac systems $\omega^{(\rm g)} \propto \sqrt{n_d}/n_d^{1/2d}$, and for parabolic systems $\omega^{(\rm p)} \propto \sqrt{n_d}$, in $d$-dimensional systems. Additionally we note that a beautiful similarity emerges between all three systems if we use the density dependent effective Dirac mass (or cyclotron mass)  for GDP and MDP,  $m_{\rm d} = \varepsilon_{\rm F}/v_{\rm F}^2$, to express the  long wavelength plasmon dispersion for all systems in all dimensions:  $ \omega^{({\rm p},{\rm g},{\rm m})} \propto \sqrt{n_d/m_{\rm p/d}}$. 
This density dependence of the plasmon frequency may be used to distinguish between various types of systems (parabolic, GDP and MDP) and their effective dimensionality.  Alternatively, the long wavelength plasmon dispersion may be used to determine the dynamical collective mass of various Dirac systems\cite{Yoon}. 

We have also calculated the plasmon dispersion in a1d-superlattice made of nano-ribbons and layers of MDP, and find that while the $q$ dependence  is similar to that of 
parabolic superlattice, the density dependence is completely different.
Note that our results  for the 3d plasmons of MDP,  2d layers and multilayers, as well as 1d ribbons and multi ribbon arrays can be tested using electron scattering, light scattering or infrared spectroscopy for ribbons and layers made of silicene, transition metal dichalcogenides and other materials which have a dispersion similar to that of massive or gapped Dirac spectrum at low energies. 

Finally we note that we have not included in our calculation the effect of dielectric mismatch, which may lead to the dielectric constant appearing in the expression of the Coulomb interaction $\kappa$, to be wave vector dependent. This can be easily included in our calculations according to Ref.~[\onlinecite{Badalyan}].  However we believe that the long wavelength plasmon dispersion, as discussed in this article,  would not be impacted by this, even though the dielectric mismatch may change the plasmon dispersion for finite wave vectors. 

\section*{Acknowledgements}  We gratefully acknowledge funding from the INSPIRE Faculty Award by DST  (Govt. of India) (AA), and from the Faculty Initiation Grant by IIT Kanpur, India (AA). We also acknowledge financial support by NSF grant DMR-1406568 (GV).

\end{document}